\documentclass[runningheads]{svmult}

\usepackage{makeidx}   
\usepackage{graphicx}  
\usepackage{subeqnar}  
\usepackage{multicol}  
\usepackage{physprbb}  
\makeindex             

\begin{document}
\title*{TELEPENSOUTH project:\\ Measurement of the Earth gravitomagnetic field in a terrestrial
                  laboratory}
\toctitle{TELEPENSOUTH project: \protect\newline Measurement of
the Earth gravitomagnetic field in a terrestrial
                  laboratory}
%
%
\titlerunning{TELEPENSOUTH project}
%
\author{J.-F. Pascual-S\'anchez}

\authorrunning{J.-F. Pascual-S\'anchez}
%
%
\institute{Dept. Matem\'atica Aplicada Fundamental, Secci\'on
Facultad de Ciencias,\\Universidad de Valladolid,
     47005, Valladolid, Spain}

\maketitle              

\begin{abstract}
We will expose a preliminary study on the feasibility of an
experiment leading to a direct measurement of the gravitomagnetic
field generated by the rotational motion of the Earth. This
measurement would be achieved by means of an appropriate coupling
of a TELEscope and a Foucault PENdulum in a laboratory on ground,
preferably at the SOUTH pole. An experiment of this kind was
firstly proposed by Braginski, Polnarev and Thorne, 18 years ago,
but it was never re-analyzed.
\end{abstract}

\section{Introduction}
The search for measurable effects
 of a  gravitational field due to the angular
momentum of the source, within the framework of General Relativity
(GR), continues. In the weak and slow motion approximation of GR,
the gravitomagnetic part of the gravitational potential gives
rise to the Lense-Thirring effect \cite{len}. The actual
detection of this effect is entrusted both to Earth  satellites
experiments and to Earth based laboratory experiments. So far,
the only positive indirect result concerns an experiment of the
first kind, the precession of the nodes of the orbit of the LAGEOS
satellite \cite{ciufolini}.

On the other hand, in the next years the space mission Gravity
Probe B (GPB) is planned to fly, carrying gyroscopes which should
verify the Lense-Thirring precession effect directly
\cite{GPB1},\cite{GPB2}. Moreover, different possibilities
connected both with the clock effect and the gravitational Sagnac
effect have been considered \cite{mashhoon},\cite{tartaglia}.

Recently, after the completion of this work, a Earth based
laboratory experiment to test directly the quadratic terms in the
angular momentum of a gravitational potential, has been proposed
by Tartaglia (see \cite{Tarta1}). This proposal deserves further
study. However, in what follows, I will remind a different Earth
based laboratory experiment to test directly the Lense-Thirring
effect, which was firstly proposed by Braginski, Polnarev and
Thorne, 18 years ago \cite{Bra}, but never reconsidered or
re-analyzed.

\section{Gravitomagnetic Maxwell-like equations}
First at all,  we give a fast review of the well-known linear and
slow motion approximation of GR. Our starting point will be the
Einstein field equations:
\begin{equation}
R_{a b}-\frac{1}{2}g_{a b}R= -\frac{8\pi G}{c^{4}}%
T_{a b} \,\,. \label{Ein_field_equ}
\end{equation}
If the gravitational field is weak, then the metric tensor can be
approximated by
\begin{equation}
g_{a b}\simeq \eta _{a b}+h_{a b} \label{1}\,,
\end{equation}
where $\eta _{a b }$ is the flat Minkowski spacetime metric. Now
define the gravitational potentials as
\begin{equation}
\overline{h}_{a b}=h_{a b}-\frac{1}{2}\eta _{a b}\,h
\label{hbar}\,.
\end{equation}
The analogy with Maxwell-Lorentz electrodynamics can be made
explicit by writing the linear gravitational equations in terms
of first-order derivatives of the gravitational potential, i.e.,
acceleration fields. With this aim in view, we first introduce
the object
\begin{equation}\label{6}
G^{abc}={\frac{1}{4}}\left(\overline{h}\,^{ab,c}-\overline{h}\,^{ac,b}
 +  \eta^{ab}\,\overline{h}\,^{cd}_{\;\;\;\; ,d} - \eta^{ac}\,
 \overline{h}\,^{bd}_{\;\;\;\; ,d}\right)\,.
\end{equation}
and impose the four harmonic de Donder
 gauge conditions:
\begin{equation}\label{7}
\overline{h}\,^{ab}_{\;\;\;\; ,b}=0.
\end{equation}
From (\ref{6}) and (\ref{7}) reads:
\begin{equation}\label{g}
 G^{abc}={\frac{1}{4}}\left(\overline{h}\,^{ab,c}-\overline{h}\,^{ac,b}\right),
\end{equation}
and retaining
 only linear terms, one obtains the weak field equations in terms
 of the object $ G^{abc}$, in which only
 first-order derivatives of the gravitational potential occur:
\begin{equation}\label{11}
\frac{\partial G^{abc}}{\partial x^c}= -\frac{4\pi G}{c^{4}}
T^{ab}.
\end{equation}
 After defining the
gravitoelectric Newtonian scalar potential $\Phi$ and
 the gravitomagnetic vector potential $\mathbf{a}$ as
 \begin{equation}
\Phi := -\frac{c^{2}\,\overline{h}\,^{00}}{4}  \label{g1}
\end{equation}
\begin{equation}
a^{i}:=\frac{c^{2}\,\overline{h}\,^{0i}}{4}
 \qquad {\mathbf{a}}=\left(a^{1},a^{2},a^{3}\right) \,\,, \label{g2}
\end{equation}
let us introduce new symbols and substitute equations (\ref{g1})
and (\ref{g2}) into equation (\ref{g}), to get the gravitoelectric
Newtonian field $\mathbf{g}$ as
\begin{equation}
{\mathbf{g}}=-\nabla \Phi - \frac{1}{c}\frac{\partial \mathbf {a} }{%
\partial t}, \label{Maxg}
\end{equation}
where
\begin{equation}
g^{i}= c^{2}G^{00i}=-\frac{\partial \Phi }{\partial
x^{i}}-\frac{1}{c}\frac{\partial {a^{i}}}{\partial t}\,\,,
\label{clov16}
\end{equation}
and
\begin{equation}
G^{00i}=\frac{1}{4}\left(
\overline{h}\,^{00,i}-\overline{h}\,^{0i,0}\right), \label{clov15}
\end{equation}
and  to get the gravitomagnetic field $\mathbf{b}$ as
\begin{equation}\label{12}
\begin{array}{llll}
 \mathbf{b}=\nabla\wedge \mathbf{a} \,\,,
&&& c^{2}G^{0ij}=a^{i,j}-a^{j,i}\,\,,  \\[5mm] {\mathbf{b}}=(b^1,b^2,b^3),
\hspace{5mm}& b^1=c^2\,G^{023}\,\,,& b^2=c^2\,G^{031}\,\,, &
b^3=c^2\,\,G^{012}  \,\, .
\end{array}
\end{equation}
 Now, performing the first order slow motion approximation for the
energy momentum tensor we neglect quadratic terms in velocity,
i.e., neglect
 the stress part of the energy-momentum tensor.
Thus, the energy-momentum tensor will only have the components
\begin{equation}
T^{00}=\rho c^{2}  \label{clov5}
\end{equation}
and
\begin{equation}
T^{0i}=\rho cv^{i}.  \label{En_comp}
\end{equation}
Thus, when the first order effects of the motion of the sources
are taken into account, one arrives at the following {\it
gravitomagnetic} (Maxwell-like) equations \cite{Pas}:
\begin{eqnarray}
\nabla \,\mathbf{g}             & =&  -4\pi G \rho\,\,, \label{13}\\
\nabla \,\mathbf{b} &=&0\,\,, \label{14} \\
 \nabla\wedge
{\mathbf{g}} &= &-\displaystyle\frac{1}{c}\frac{\partial
\mathbf{b}}{\partial t}\,\,,
\label{15} \\
\nabla\wedge \mathbf{b} &=& -\frac{4\pi G }{c}\,\rho\,
{\mathbf{v}}+ \displaystyle\frac{1}{c} \frac{\partial
{\mathbf{g}}}{\partial t}\,\,. \label{16}
\end{eqnarray}
For a weak stationary gravity field, from the geodesic equation

\begin{equation}\label{18}
\frac{d^2x^a}{d\tau^2}+\Gamma^a_{bc}\,\frac{dx^b}{d\tau}\frac{dx^c}{d\tau}=0\,\,,
\end{equation}
 one obtains the Lorentz-like force law, reads
\begin{equation}\label{19}
\frac{d\mathbf{u}}{dt} ={\mathbf{g}}+ \frac{4}{c}
\,\mathbf{u}\wedge \mathbf{b} ,
\end{equation}
where $\mathbf{u}$ is the velocity of the test particle.

\section{Lense-Thirring precession on a spin}
For a weak stationary field  the gravitomagnetic potential is
\begin{equation}\label{20}
{\mathbf{a}} =-{\frac{1}{2}}\,\frac{G}{c}\,\frac{\mathbf{J}
\wedge \mathbf{r}}{r^3},
\end{equation}
and the gravitomagnetic field
\begin{equation}\label{21}
{\mathbf{b}}=\nabla\wedge
{\mathbf{a}}=-{\frac{1}{2}}\,\frac{G}{c}\,\frac{3\mathbf{n}\,(\mathbf{J}\cdot
\mathbf{n})-\mathbf{J}}{
 r^3}
\end{equation}
where $\mathbf{J}$ is the intrinsic angular momentum of the source
and $\mathbf{n}$ is the unit position vector. These equations are
analogous to the magnetostatic ones, replacing the magnetic dipole
moment by minus half the angular momentum. \\
 For an arbitrary accelerated observer in a
gravitational field with 4-velocity $u^{a}=dx^{a}/d\tau$ and
4-acceleration $a^{a}=Du^{a}/d\tau$, the equation of motion for
the torque-free point-like 4-spin vector is given by the
Fermi-Walker transport law
\begin{equation}
\frac{dS^{a}}{d\tau}+\Gamma^{a}_{b\,c}\hspace{.05in}u^{b}S^{c}=u^{a}a_{d}S^{d},
\label{spin}
\end{equation}\\
where the 4-spin vector $S^{a}$ is constrained by the condition
\begin{equation}
u_{a}S^{a}=0 \,,
\end{equation}
 which assures that the length of the 3-spin
vector $\mathbf{S}$ does not change as measured by an observer
comoving with the spinning particle. The total precession of the
3-spin vector with respect to an asymptotic inertial frame given
by a ``fixed star" trained on by a telescope, whose associated
tetrad  realizes Frenet-Serret transport, is given by the equation
\begin{equation}
\frac{d\mathbf{S}}{dt} =\,\mathbf{\Omega}\,\wedge\,\mathbf{S} .
\end{equation}

The general expression for the spin precession rate in the
Lense-Thirring metric (Schiff formula) contains three terms
\begin{equation}
\mathbf{\Omega} =
\mathbf{\Omega_{Th}}+\mathbf{\Omega_{geo}}+\mathbf{\Omega_{LT}}\,\,,
\end{equation}
where
\begin{equation}
{\mathbf{\Omega_{Th}}}= \frac{1}{2c^2} \,\mathbf{a} \wedge
\mathbf{u}\,\,,
\end{equation}
and
\begin{equation}
{\mathbf{\Omega_{geo}}}=
\frac{3}{2}\frac{G}{c^2}\frac{M}{r^2}\,{\mathbf{n}\wedge
\mathbf{u}}=\frac{3}{2\,c^2}\mathbf{u}\wedge \mathbf{g}\,.
\end{equation}\\
Only the Thomas precession $\mathbf{\Omega_{Th}}$ would be present
for accelerated motion in a flat Minkowski spacetime. Both
geodetic $\mathbf{\Omega_{geo}}$ and Thomas
$\mathbf{\Omega_{Th}}$ precessions
 are present for accelerated motion in
Schwarzschild geometry, but only the geodetic
$\mathbf{\Omega_{geo}}$ remains, in the case of free fall
($\mathbf{a =0}$) motion. Moreover, the geodetic de Sitter-Fokker
precession $\mathbf{\Omega_{geo}}$, due to the mass $M$, is in
the same sense as the
orbital motion.\\

 The additional Lense-Thirring gravitomagnetic
precession effect $\mathbf{\Omega_{LT}}$, due to the angular
momentum $\mathbf{J}$ of the source, is  manifested in a Kerr
spacetime \cite{Ura} or in its weak field and slow motion
approximation (the Lense-Thirring metric \cite{Sch}), in which the
Lense-Thirring precession rate $\mathbf{\Omega_{LT}}$ is

\begin{equation}
{\mathbf{\Omega_{LT}}} = - \frac{2}{c} \,\,{\mathbf{b}} =
\frac{G}{c^2}\,\,\frac{3\mathbf{n}\,(\mathbf{J}\cdot
\mathbf{n})-\mathbf{J}}{
 r^3}
\end{equation}

\subsection{Particular cases of the Schiff formula}
\emph{(1) Free fall gyroscopes in Earth's orbit (GP-B experiment).}\\
 As $\mathbf{a}=\mathbf{0}$ and $\mathbf{u}
 \neq \mathbf{0}$, then $\mathbf{\Omega_{Th}=0}$ and only two terms survive,
\begin{equation}
\mathbf{\Omega} = \mathbf{\Omega_{geo}}+\mathbf{\Omega_{LT}}
\end{equation}
 This formula is used in the GP-B
gyroscope (Stanford) experiment to obtain the precession, due to
stationary gravitomagnetic $\mathbf{b}$ field generated by the
rotation of the Earth mass, with respect to an asymptotic
inertial frame given by the ``fixed star" TM Pegasus.\\

\noindent \emph{(2) Gyroscope at rest on Earth (except at the poles).}\\
In this case, $\mathbf{a}= - \mathbf{g}$, hence
\begin{equation}
\mathbf{\Omega} =
\mathbf{\Omega_{Th}}+\mathbf{\Omega_{geo}}+\mathbf{\Omega_{LT}}
\end{equation}
with $\mathbf{\Omega_{Th}}+ \mathbf{\Omega_{geo}}=
\frac{2}{c^2}\mathbf{u}\wedge\mathbf{g}$\,\,,
\\

\noindent \emph{(3) Gyroscope or Foucault pendulum at rest on
Earth at a pole (South).}
\\
As $\mathbf{u=0}$, then only the Lense-Thirring term remains
\begin{equation}
\mathbf{\Omega} =\mathbf{\Omega_{LT}}\,\,.
\end{equation}
Hence, this last is the clean experiment because one has the GM
precession only, without competing geodetic or Thomas effects.
Moreover, the magnitude of $\mathbf{\Omega_{LT}}$ is five times
larger in this last case ($220\,\,mas/year$) than in the GP-B
experiment ($42\,\,mas/year)$.

\section{TELEPENSOUTH experimental apparatus}
The experimental apparatus would consist of a Foucault pendulum
and an astrometric telescope in a underground vacuum chamber. To
avoid the classical Foucault effect, due to Earth's rotation, it
is necessary to operate just in the South pole.

Furthermore, another reason for using a Foucault pendulum instead
of a gyroscope is due to the required sensitivity compared to the
Earth rotation rate: $\Omega_{LT}= d\Phi/d\,t\approx
\,5.10^{-10}\,\, \omega_\oplus$.

The telescope must have its optical axis locked to the azimuth of
a ``fixed" star, which is an approximate asymptotic inertial
frame.

The pendulum swinging fiber is used as a light pipe and the mass
as a lens to focus a swinging light beam onto an optical system,
which monitors the angle $\Phi$ between the principal axis of the
pendulum and the telescope and the ellipticity of the swing
$\epsilon$.

\subsection{Sources of error of the experiment}
For the sake of completeness, we give here a summary of the
sources of error of the experiment, which has been mainly
extracted  from the original work \cite{Bra}. For a two-month
experiment, $10\,\%$ of accuracy requires a precision of
$\delta\Phi = 4\,\, mas$.

\vspace{.2cm}\noindent \emph{(A) For the pendulum, two different
kinds of error}:

(1) Velocity dependent forces: Gravitomagnetic (to measure),
magnetic, aniso-tropic
 frictional damping, Pippard precession.

 \noindent Effect: Change of the principal axis direction $\Phi$,
no change in $\epsilon$.

(2) Position dependent forces orthogonal to the principal axis of
the pendulum: frequency anisotropy, seismic displacements of the
support.

 \noindent Effect: First order change in $\epsilon$, second
order in $\Phi$.

\vspace{.2cm} \noindent \emph{(B) For the telescope, the errors
come from}:

 (1) Atmospheric refraction.

  (2) Distortion.

(3) Tilts of the mirror.

\subsection{Control of the sources of error of the pendulum}
(1) Velocity dependent forces:

(a) Against magnetic forces: Coat the mass and the fiber with
metal.

(b) Against anisotropic frictional damping: pendulum support held
fixed relative to the Earth. Conclusion: Sapphire fiber with
diameter $d= 0.1\,\,mm$ and mass, $M = 100\,\, gr$.

(c) Pippard precession: due to the spin of the pendulum mass, if
the support is fixed relative to the Earth. It is $10^3$ times
larger than the GM effect. Remedy: subtract it from the data,
mass must be long, thin and dense, e.g., tungsten.

\vspace{.2cm} \noindent (2) Position dependent forces:

(a) Frequency anisotropy: due to the ellipticity $\epsilon$ and
large amplitude $A = 5\,\, cm$. Cure: Length of the fiber, $l
=2\,\, m$, and gravitational and electrostatic pulls of large
masses in parallel plates placed on each side of the pendulum or
use of  a pendulum without fiber, with the magnetically levitated
mass sliding over a superconducting surface.

(b) Seismic noise. Remedy: sapphire fiber.

\section{Acknowledgements}
I am grateful to A. San Miguel and F. Vicente for discussions on
this topic. Also, I thank L. Flor\'{\i}a for a reading of the
manuscript. This work is partially supported by the project
VA014/02 of the Junta de Castilla y Le\'on.

%

\end{document}